\documentclass[10pt]{IEEEtran}

\usepackage{amsmath}
\usepackage{mathtools}
\usepackage{graphicx}
\usepackage{enumerate}
\usepackage{url}
\usepackage{subfigure}
\usepackage{algorithm}
\usepackage{algpseudocode}
\usepackage{stfloats}
\usepackage{lineno}
\usepackage{hyperref}
\usepackage{bm}
\usepackage{bbm}
\usepackage{amssymb}
\usepackage{enumerate}
\usepackage{xcolor}
\usepackage{float}
\usepackage{cite}
\usepackage{tabularx}
\usepackage{tabu}
\usepackage{multicol}
\usepackage{multirow}
\usepackage{colortbl,booktabs,threeparttable}
\usepackage{dcolumn}
\usepackage{flushend}
\usepackage{soul}
\usepackage{CJKutf8}
\usepackage{ragged2e}

\graphicspath{{Figures/}}

\newcommand{\rscl}[1]{\mathrm{#1}}  % random scalar
\renewcommand{\vec}[1]{\bm #1}
\newcommand{\rvec}[1]{\mathbf{#1}}
\newcommand{\mat}[1]{\bm #1}

\newcommand{\bb}[1]{\mathbb{#1}}
\renewcommand{\cal}[1]{\mathcal{#1}}

\newcommand{\E}{\mathbb{E}}

\renewcommand{\P}{\mathbb{P}}

\renewcommand{\H}{\mathsf{H}}

\definecolor{hl-bg-color}{RGB}{255,255,215}
\sethlcolor{hl-bg-color} 
\definecolor{new-magenta}{RGB}{255,0,255}
\soulregister\cite7
\soulregister\citep7
\soulregister\citet7
\soulregister\ref7
\soulregister\eqref7
\ifdefined\HIGHLIGHT

\else

\fi

\newtheorem{example}{{Example}}

\begin{document}
\newpage
\title{Machine Learning in Communications: A Road to Intelligent Transmission and Processing}
%\title{First Ten Years of Machine Learning in Communications}

\author{Shixiong Wang %,
       % Wei Dai,
        and Geoffrey Ye Li,~\IEEEmembership{Fellow,~IEEE}% <-this % stops a space
\thanks{Shixiong Wang and Geoffrey Ye Li are with the Intelligent Transmission and Processing Laboratory (ITP Lab), Department of Electrical and Electronic Engineering, Imperial College London, London SW7 2AZ, United Kingdom (E-mails: s.wang@u.nus.edu; geoffrey.li@imperial.ac.uk).
%(\textit{Corresponding Author: S. Wang.})
}
}

\maketitle

% abstract
\begin{abstract}
Prior to the era of artificial intelligence and big data, wireless communications primarily followed a conventional research route involving problem analysis, model building and calibration, algorithm design and tuning, and holistic and empirical verification. However, this methodology often encountered limitations when dealing with large-scale and complex problems and managing dynamic and massive data, resulting in inefficiencies and limited performance of traditional communication systems and methods. As such, wireless communications have embraced the revolutionary impact of artificial intelligence and machine learning, giving birth to more adaptive, efficient, and intelligent systems and algorithms. This technological shift opens a road to intelligent information transmission and processing. This overview article discusses the typical roles of machine learning in intelligent wireless communications, as well as its features, challenges, and practical considerations.

% \begin{figure}[!htbp]
%     \centering
%     \includegraphics[width=7cm]{Figures/fig_role_of_ML.pdf}
%     \caption{Roles of machine learning in intelligence transmission and processing are to mathematically model complex problems and algorithmically solve complex models. (Icon Credit: FLATICON.com.)}
%     \label{fig:concept-intelligence-2}
% \end{figure}
\end{abstract}

% keywords

\section{Introduction} \label{sec:introdction}
Since the 19th century, radio communications have started a new era of information transmission for human society. The early stages of radio transmission technology relied heavily on manual operations, such as Morse codes and telegraph machines, which limited the efficiency and reliability of information exchange. Aiming to automate the task of information transmission and processing at a sophisticated level, the first-generation concept of ``intelligent transmission and processing" emerged. Subsequently, the 20th century witnessed significant developments in module-based communication systems. These systems encompass essential modules, such as source coding, channel coding, modulation, transmit beamforming, wireless channel transmission, receive beamforming, demodulation, signal detection, channel decoding, and source decoding \cite{tse2005fundamentals,cover2006elements,stuber2017principles}. Methodologically, the development of module-based wireless communications and signal processing follows a systematic research trajectory that includes problem analysis, model development and calibration, algorithm design and optimization, and empirical validation, feedback, and improvement. Notably, every step in this methodological loop demands large volumes of human intellectual endeavors.

In the 21st century, wireless communication systems are expected to deliver extremely vast amounts of data in various formats, such as audio, video, and text, while ensuring low latency, high data rates, and reliability. Furthermore, the incorporation of new network topologies (e.g., internet-of-things networks, unmanned-aerial-vehicle relay networks) and cutting-edge functions (e.g., integrated sensing and communications, integrated computing and communications) has added complexity to the design of modern communication systems. This complexity is particularly evident in the following three aspects: 
\begin{enumerate}
    \item addressing different types of modeling uncertainties in not only holistic systems but also individual modules; 
    \item leveraging various forms of big data generated by user equipment and base stations; 
    \item solving challenging algorithmic problems in realizing the networks. 
\end{enumerate}
Traditional design methods rely heavily on the intensive intellectual efforts of humans and are proven inadequate for managing large-scale and complex issues and handling dynamic extensive data. This inadequacy results in inefficiencies and limited performance of information transmission and processing. In response, wireless communications and signal processing have embraced the transformative potential of artificial intelligence (AI) and machine learning (ML) \cite{tong2022nine}; for comprehensive surveys of machine learning on communications, see, e.g., \cite{jiang2016machine,qin2019deep,gunduz2019machine,wang2020thirty,yu2022role,alhammadi2024artificial,celik2024dawn}. This technological and methodological shift has enabled the development of more adaptive, efficient, robust, and intelligent systems and algorithms. Consequently, the second-generation concept of ``intelligent transmission and processing" is emerging, aiming to significantly reduce the need for human intellectual efforts and improve the integrated performances of communication systems. 

\begin{figure}[!htbp]
    \centering
    \subfigure[Free of operation]{
        \includegraphics[height=3.2cm,width=3.2cm]{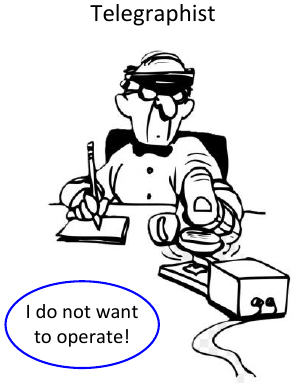}
    }~~~~~~~~~~~~~~
    \subfigure[Free of thinking]{
        \includegraphics[height=3.2cm,width=2.3cm]{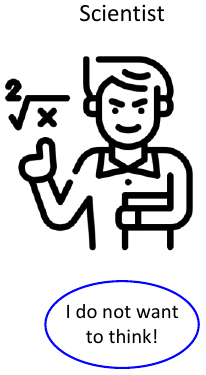}
    }
    \caption{Connotations of intelligent transmission and processing. (Icon Credit: CLEANPNG.com and FLATICON.com.)}
    \label{fig:concept-intelligence}
\end{figure}

\begin{figure}[!htbp]
    \centering
    \includegraphics[height=3cm]{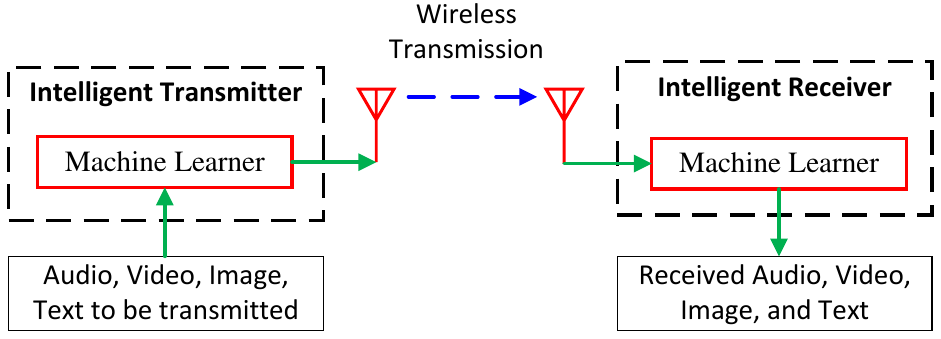}
    \caption{An end-to-end structure of intelligent transmission and processing systems. The intelligent transmitter and receiver act as end-to-end information processors $f$, where $f$ are learned by machines from data; the transmitter and receiver can automatically adapt to the real-time characteristics of wireless channels. The intelligence is reflected in the sense that human intellectual efforts are no longer explicitly required to study large-scale, dynamic, and uncertain information transmission mechanisms and processing solutions.}
    \label{fig:intelligent-transceiver}
\end{figure}

Fig. \ref{fig:concept-intelligence} illustrates the philosophical connotations of intelligent transmission and processing. A technical visualization of ML-empowered intelligent transmission and processing is shown in Fig. \ref{fig:intelligent-transceiver}.

To showcase the power of ML techniques in enabling intelligent transmission and processing, this article reviews trending ML applications in communication systems and methods, including physical-layer communications \cite{qin2019deep,he2019model}, semantic communications \cite{xie2021deep,jiang2023wireless}, resource allocations in communications \cite{liang2019deep}, integrated computing and communications \cite{ferrag2023edge}, and integrated sensing and communications \cite{lu2024integrated}. However, the ambition of this article is not to offer an exhaustive list of all existing works in the area. Rather, we aim to illuminate the path for intelligent transmission and processing.

\begin{figure}[!htbp]
    \centering
    \includegraphics[width=7.5cm]{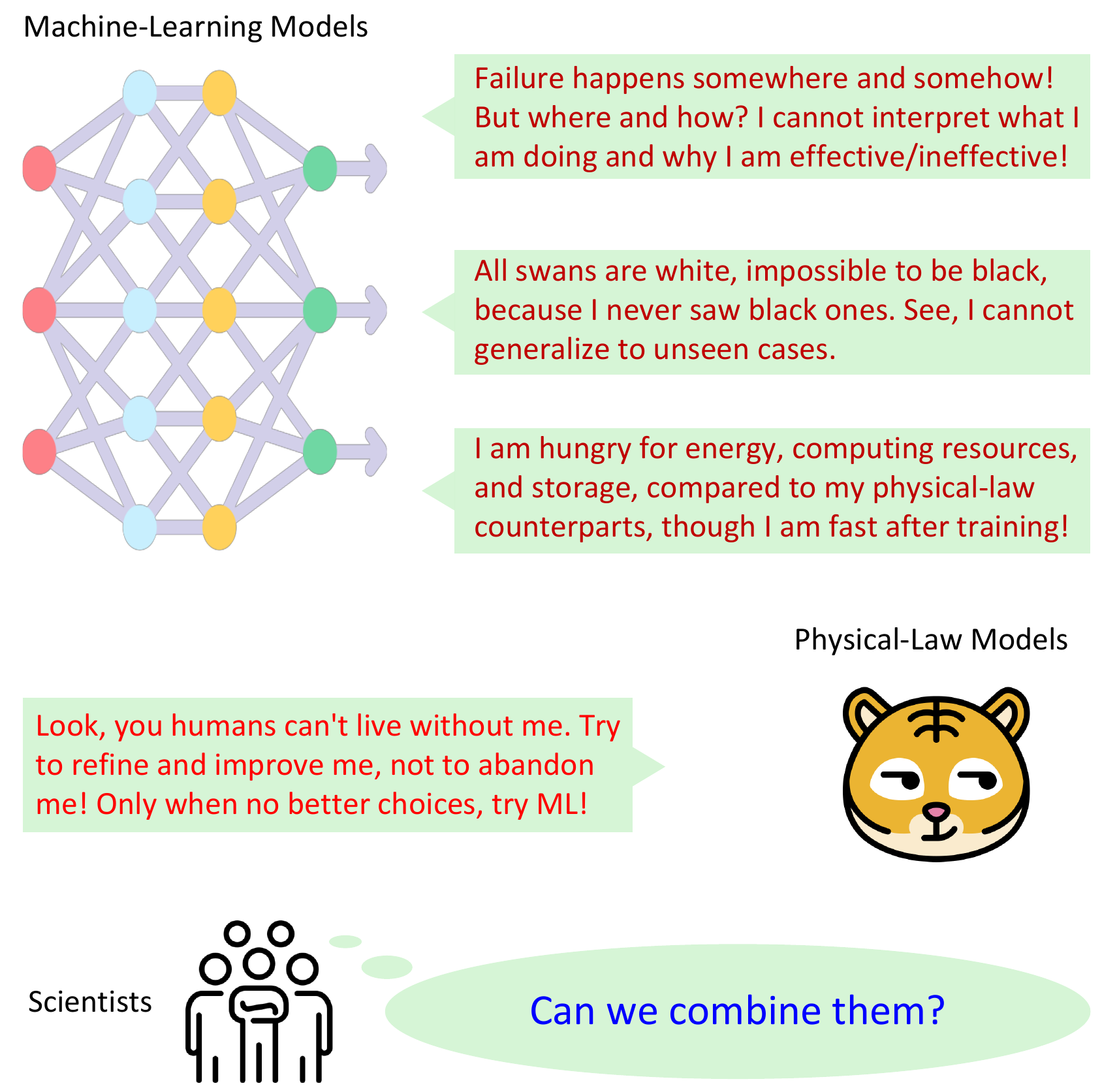}
    \caption{To choose a machine-learning model or a physical-law model, this is a question! Nothing is free, although some are cheap! Choose the one that fits your situation and expectations well! Whenever possible, combine them to improve the overall system performance. (Icon Credit: FLATICON.com.)}
    \label{fig:nothing-free}
\end{figure}

Although the role of machine learning is promising in reforming the theory and practice of wireless communications, the challenges and disadvantages of utilizing ML-based approaches accompany its opportunities and advantages \cite{tong2022nine}, for example, the reliability issue due to the lack of interpretability of black-box learning methods (e.g., deep learning), the generalization issue due to the limited training data and the non-stationarities of the underlying data-generating laws, and the resource deficits in training and storing large ML models (e.g., deep learning); see Fig. \ref{fig:nothing-free} for a motivational understanding. In addition to the three primary challenges exemplified, other instances may also arise, e.g., the scalability issue caused during the reconfiguration of system topology or hardware (e.g., removing or adding antennas; which can be seen as a kind of generalization problem) and the security and privacy issue in networked learning \cite{ferrag2023edge}. The main message is that in advancing communication theories and developing communication systems, the role of machine learning should not be overstated: machine learning (especially data-driven deep learning) can be a valuable factor to consider rather than an absolute rule to follow; problem analyses and mechanism modeling are always important; see, e.g., \cite{monga2021algorithm,shlezinger2023discriminative,wang2024distributionally} for technical investigations and justifications; see also the example below for a motivational understanding.
\begin{example}[Ice Cream Sales and Shark Attacks]
Regression analysis using historical data shows a positive relationship between ice cream sales and shark attacks, which is non-logical. However, the primary factor driving this correlation is temperature: higher temperatures lead to increased ice cream sales and beach attendance; more beach visitors result in more shark attacks \cite[p.~75]{james2021introduction}. Hence, mechanism modeling is vital.
\end{example}

Before digging into ML applications in communications, we quickly review the essentials of ML concepts and methods in Section \ref{sec:ML-theory}, especially those of trustworthy machine learning. The aim is to highlight the primary considerations, including philosophical and technical facets, of using machine learning in wireless communications.

\section{ML Concepts and Methods}\label{sec:ML-theory}
Machine learning is concerned with discovering hidden information and patterns from data. The primary advantage is its ability to explain data automatically, thus liberating humans from studying the underlying data-generating mechanisms. This feature inherently enables machine intelligence in the practice of communications, specifically, in the transmission and processing of information \cite{jiang2016machine,sun2019application,gunduz2019machine,wang2020thirty,celik2024dawn}. 

\begin{figure}[!htbp]
    \centering
    \includegraphics[height=5cm]{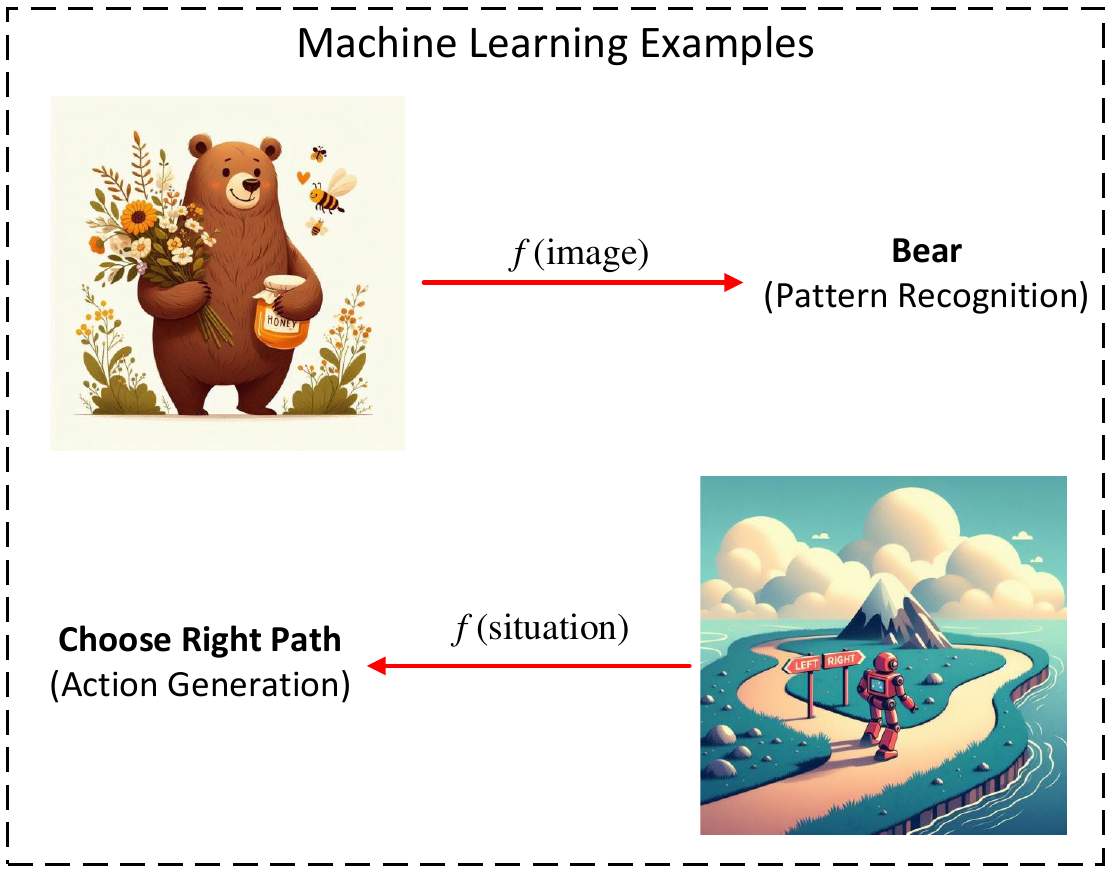}
    \caption{Conceptual illustration of machine learning. Machine learning is to find a mapping $f$ from the input data to a decision. The upper example is a supervised machine learning problem where an image classifier $f$ recognizes the image as a bear. The lower example is a reinforcement learning problem where an action generator $f$ recommends the robot to choose the right path in the current situation. (The two images are generated by Microsoft Copilot.)}
    \label{fig:concept-machine-learning}
\end{figure}

Depending on the characteristics of tasks, machine learning can be categorized into four genres: supervised learning, unsupervised learning, semi-supervised learning, and reinforcement learning. Mathematically, the key to all machine learning tasks is to find a function $f$, called a hypothesis, that maps the observed data to a desired decision; see Fig. \ref{fig:concept-machine-learning} for a conceptual illustration.
Specific examples of machine learning are as follows.
\begin{enumerate}
    \item \textbf{Supervised Learning}: Supervised learning summarizes the hidden information of labeled data and includes regression and classification as two main types of tasks. Given the deterministic or random variable pair $(\rvec x, \rvec s)$ where $\rvec x$ is the feature vector and $\rvec s$ is the continuous-valued expected response, \textbf{regression} aims to find a functional relation $f$ from $\rvec x$ to $\rvec s$ such that predicted label $f(\rvec x)$ can be as close as possible to the target label $\rvec s$. The closeness is measured by a loss function $L$ through $L[\rvec s, f(\rvec x)]$, for example, the mean-squared error $[\rvec s - f(\rvec x)]^\H [\rvec s - f(\rvec x)]$. In handling deterministic variable pair $(\rvec x, \rvec s)$, there potentially exists a function $f$ such that $f(\rvec x)$ is exactly equal to $\rvec s$ for every realization of $(\rvec x, \rvec s)$, i.e., $L[\rvec s, f(\rvec x)] \equiv 0$. As for random variable pair $(\rvec x, \rvec s)$, however, the exact equality cannot be generally guaranteed. Instead, the loss is calculated under the joint distribution $\P_{\rvec x, \rvec s}$ of $(\rvec x, \rvec s)$, for example, the expectation of the loss $\E_{(\rvec x, \rvec s) \sim \P_{\rvec x, \rvec s}}L[\rvec s, f(\rvec x)]$. When $\rscl s$ is one-dimensional and takes discrete values,  we have the \textbf{classification} problems where $L[\rscl s, f(\rvec x)]$ is defined by, e.g., the indicator function $\bb I\{\rscl s \neq f(\rvec x)\}$. In this case, $\rscl s$ indexes different target classes; e.g., for binary classification, $\rscl s \in \{-1, 1\}$.

    \item \textbf{Unsupervised Learning}: In unsupervised learning, there are no labels $\rvec s$ for the collected data, and only feature data $\rvec x$ are present. Therefore, we focus on discovering the hidden information from the realizations of datum variable $\rvec x$. \textbf{Clustering} is a typical task of unsupervised learning. Different from classification which predicts the categorical labels $\rscl s$ of new data points $\rvec x$ based on a training dataset with known labels, clustering aims at grouping similar data points $\rvec x$ together based on their features without predefined categorical labels. In summary, clustering is to find a function $f$ that maps data $\rvec x$ into a suitable group. \textbf{Feature transformation} is another example of unsupervised learning, which transforms original feature data $\rvec x$ into another feature space using a learned mapping $f$; i.e. $\rvec y = f(\rvec x)$; cf. a time-domain signal $\rvec x$ and its Fourier transform $\rvec y$. Autoencoders, a type of artificial neural network, provide an excellent example of feature transformation through their encoding and decoding operations. Yet another important unsupervised learning is \textbf{distribution estimation}, i.e., to estimate the data-generating distribution that best fits (or describes) the collected data. Distribution estimation is particularly vital in generative tasks such as producing a new sample based on collected samples; for example, given a group of cat images, how to produce a new cat image by drawing from the fitted distribution?

    \item \textbf{Semi-supervised Learning}: Semi-supervised learning can be considered a variant of supervised learning as it extends the principles of supervised learning by incorporating a mixture of labeled data $(\rvec x, \rvec s)$ and unlabeled data $\rvec x^\prime$. While supervised learning relies entirely on labeled data to train the model, semi-supervised learning aims to improve model performance and generalization capability by leveraging the additional unlabeled data. The semi-supervised approach is particularly useful when acquiring a large amount of labeled data is expensive or time-consuming, while unlabeled data is abundant and easy to obtain. By leveraging the information from the unlabeled data along with the labeled data, semi-supervised learning can find a model $f$ that has better prediction performance (of the label $\rvec s$ associated with the data $\rvec x$) compared to purely supervised learning methods that rely solely on labeled data.

    \item \textbf{Reinforcement Learning}: Reinforcement learning is concerned with decision-making problems in a dynamic and uncertain environment. Unlike supervised learning, which uses labeled data, and unsupervised learning, which finds patterns in unlabeled data, reinforcement learning involves the agent interacting with the environment, receiving feedback in the form of rewards or penalties, and using this feedback to learn optimal behaviors or strategies over time. To be specific, an agent autonomously learns to make decisions in an environment by performing actions $\rvec a$, in response to current states $\rvec s$, to maximize cumulative reward. Therefore, mathematically, an action-generating function $f$ from state $\rvec s$ to action $\rvec a$ needs to be learned.
\end{enumerate}
For specific applications of the four machine learning diagrams in wireless communications, refer to, e.g., \cite{wang2020thirty,eldar2022machine}.

\textbf{Data-Driven and Model-Driven Learning}: Considering the degree of human intelligence and domain knowledge involved, machine learning can be classified into data-driven and model-driven approaches. Data-driven machine learning relies entirely on historical data and no analyses on underlying data-generating mechanisms are conducted. In contrast, model-driven machine learning incorporates, to differing extents, studying the underlying physical mechanisms and data-generating models. Intelligent information transmission and processing can benefit, in terms of improving overall performance, from the collaboration between communication-systems modeling and big data discovery \cite{he2019model}, \cite[Chapter~6]{eldar2022machine}. For example, in signal detection, suppose that we have $T$ pilot data pairs $\{(\vec s_1, \vec x_1),(\vec s_2, \vec x_2),\ldots,(\vec s_T, \vec x_T)\}$ where $\vec x_i$ are the received signals and $\vec s_i$ are the transmitted symbols, for $i = 1,2,\ldots,T$. Data-driven machine learning directly utilizes all the data to train a detector $f$ from $\rvec x$ to $\rvec s$. In contrast, model-driven machine learning first considers the signal-transmission model $\rvec x = \mat H \rvec s + \rvec v$ where $\mat H$ denotes the channel matrix and $\rvec v$ the channel noise, and then finds a detector $f$ based on the above underlying data-generating mechanism. For detailed technical treatments and discussions, see, e.g., \cite{shlezinger2023model,shlezinger2023discriminative,he2020model,wang2024distributionally}.

\textbf{Hypothesis Space and Deep Learning}: To locate a best decision function $f$, a candidate function space $\cal H$ from which $f$ is drawn, called hypothesis space, needs to be specified. To clarify further, for instance, supervised statistical machine learning can be formulated as 
\[
   \displaystyle \min_{f \in \cal H}  \E_{(\rvec x, \rvec s) \sim \P_{\rvec x, \rvec s}} L (\rvec s, f(\rvec x)),
\]
where the joint distribution $\P_{\rvec x, \rvec s}$, which is unknown in practice, can be estimated using collected historical data (e.g., using empirical distribution). As an example, signal detection problems can be characterized as above, where $f$ is a detector, $\rvec x$ is the antenna-received signal, and $\rvec s$ is the transmitted symbol (e.g., constellation points) \cite{wang2024distributionally}; the loss function $L$ can be mean-squared error or symbol-error rate. Canonical examples for hypothesis space $\cal H$ can be as follows.
\begin{itemize}
    \item Linear Function Space: To be specific, $\cal H$ only includes the linear transforms of input $\rvec x$. In the signal detection case, $\cal H$ contains only linear detectors.

    \item Reproducing Kernel Hilbert Space: To be specific, $\cal H$ includes all linear transforms of the nonlinearly-lifted-feature $\varphi(\rvec x)$ of the original feature $\rvec x$, using some feature mapping function $\varphi$. In essence, $\cal H$ contains some specific types of nonlinear functions of input $\rvec x$.

    \item Neural Network Function Space: To be specific, $\cal H$ is represented (or structured, characterized) by neural networks, for instance, multi-layer perceptron, recurrent neural networks, convolutional neural networks, radial basis neural networks, autoencoders, transformers, among many others. Each given neural network defines a special type of function space $\cal H$. When the employed neural network has deep structures with many hidden layers being included, $\cal H$ denotes a deep-neural-network function space. Upon operating with deep neural networks, machine learning is referred to as \textbf{deep learning}.
\end{itemize}
On the other hand, with the involvement of domain knowledge and expert designs, a hypothesis space $\cal H$ can be accordingly adapted or tailored to a domain-specific problem \cite{he2019model,monga2021algorithm}, \cite[Chapter~6]{eldar2022machine}. Therefore, model-driven machine learning is to devise an ad-hoc and structured candidate space $\cal H$, by leveraging known problem characteristics and data-generating mechanisms.

\textbf{Explainability, Reliability, and Sustainability}: Modern machine learning research addresses several advanced concerns, including explainability, reliability, and sustainability of learning models \cite{thuraisingham2022trustworthy,Varshney2022}. Explainable machine learning seeks to make learning models transparent, interpretable, and accountable through techniques such as feature engineering and physical modeling \cite{christoph2020interpretable}; model-driven machine learning, which leverages underlying physical data-generating mechanisms, can be seen as such a scheme \cite{he2019model,monga2021algorithm}. Reliable machine learning focuses on creating robust and accurate learning models that generalize well to new data (that are not used in the training stage), tackling issues such as overfitting, generalization, knowledge migration, and limited-sample learning \cite{kawaguchi2022robustness,wang2022generalizing,wang2024distributionally,wang2024distributional}. Sustainable machine learning aims to develop learning models with minimal negative impact on the environment and society, addressing energy efficiency, privacy and security, and fairness and bias \cite{van2021sustainable,ferrag2023edge}. In the context of intelligent information transmission and processing, the three considerations (i.e., explainability, reliability, and sustainability) are of natural importance and significance. Therefore, they are the primary considerations in developing machine-learning-based solutions for wireless communications.

\textbf{Centralized and Distributed Learning}: Machine learning models can be trained using various approaches depending on the structure of the data distribution and the architecture of the computation. Two primary paradigms in this context are centralized learning and distributed learning \cite{abdulrahman2020survey,chen2021distributed}. Centralized learning involves collecting and storing all training data in a single central location, such as a data center or cloud server, and the machine learning model is trained on this aggregated dataset. Distributed learning, on the other hand, involves training machine learning models in a distributed manner across multiple devices (or nodes), each of which holds a portion of the data. \textbf{Federated learning} is a prominent example of distributed learning, where multiple clients (e.g., smartphones, internet-of-things devices, or different organizations) collaboratively train a model without sharing their local data. Instead, each client trains the model on its local data and only shares the model updates (gradients or weights) with a central server, which aggregates these updates to form a global model. Both centralized and distributed learning methods are beneficial for advancing future-generation communication systems because they can adapt to diverse modern communication network typologies.

\section{Physical Layer Communications}\label{sec:PHY}
\begin{figure}[!htbp]
    \centering
    \includegraphics[height=3cm]{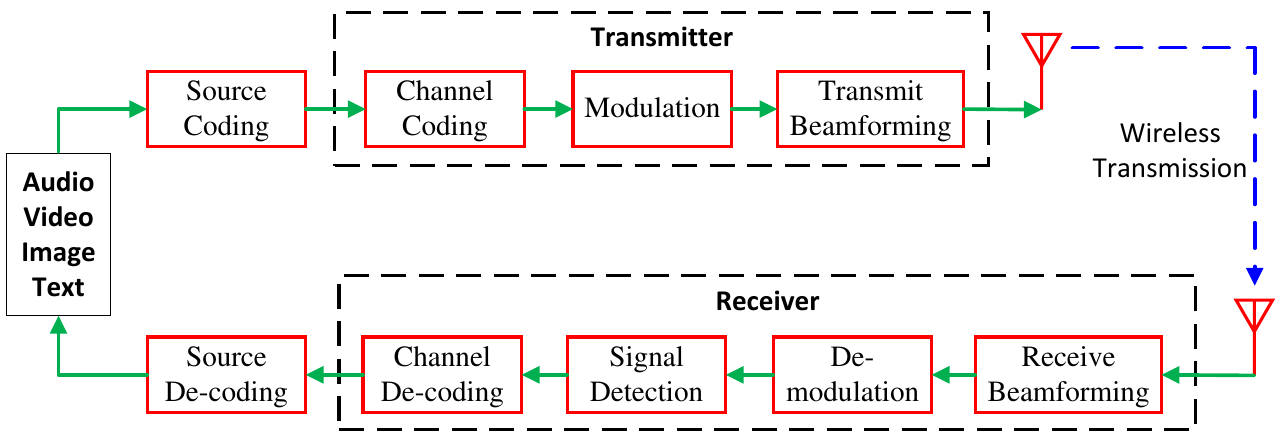}
    \caption{The module-based structure of traditional transmission and processing systems. Every block acts as an information processor $f$, where $f$ is elaborately designed by scientists based on underlying physical mechanisms and mathematical laws.}
    \label{fig:traditional-transceiver}
\end{figure}

Physical layer communications aim to reliably transmit raw data streams, e.g., binary bits, through physical mediums. Fig. \ref{fig:traditional-transceiver} presents a traditional architectural diagram of wireless communications, featuring various functional modules (or blocks) that are meticulously designed by humans in accordance with fundamental mathematical and physical principles. This block-based diagram is structurally different from the ML-empowered architectural diagram in Fig. \ref{fig:intelligent-transceiver}, where interconnected functional modules are taken over by end-to-end operating parts.

\begin{figure}[!htbp]
    \centering
    \includegraphics[height=5.5cm]{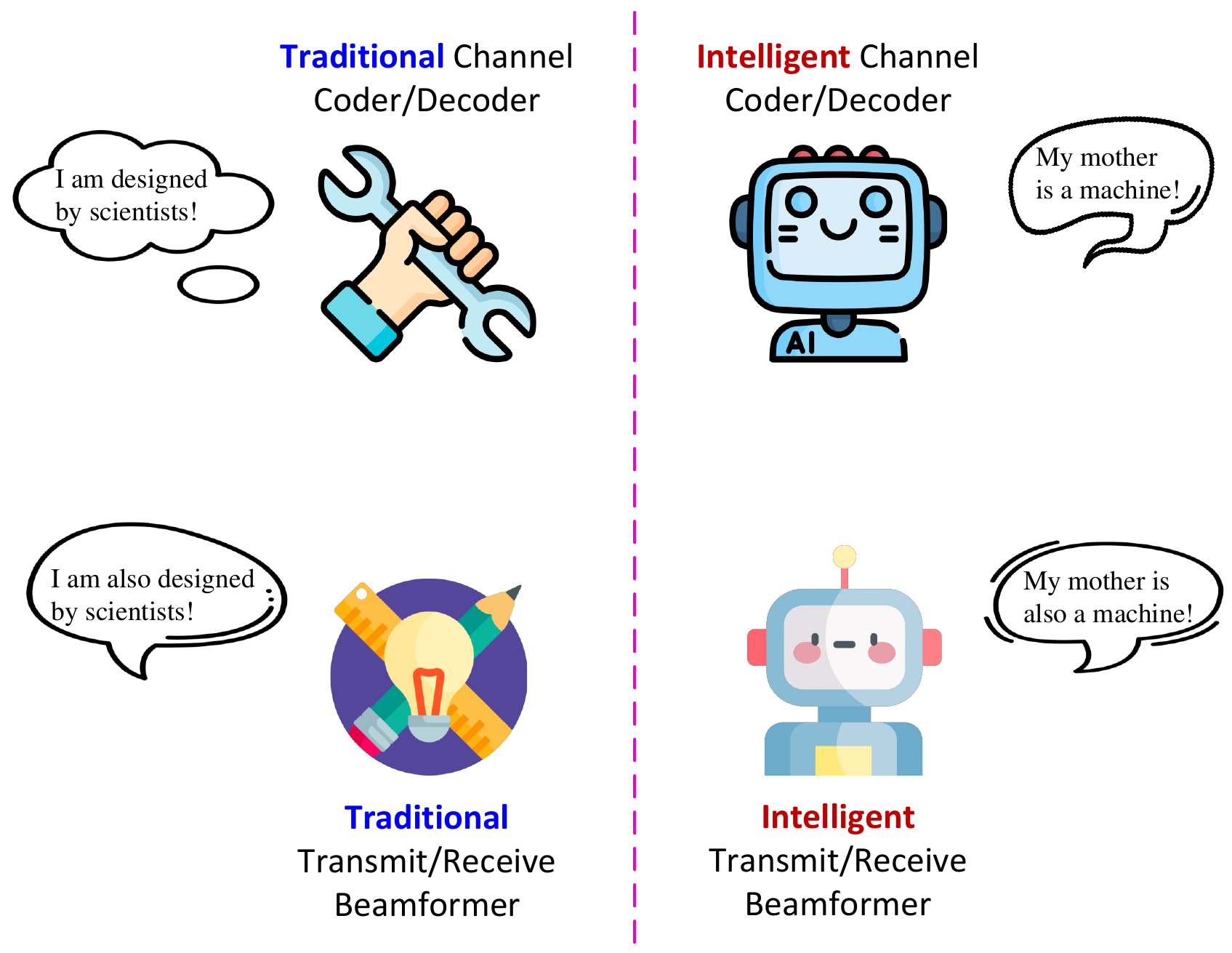}
    \caption{In contrast to the highly integrated structure in Fig. \ref{fig:intelligent-transceiver}, every individual module in Fig. \ref{fig:traditional-transceiver} can be augmented by machine learning. (Icon Credit: FLATICON.com.)}
    \label{fig:intelligent-modules}
\end{figure}

In addition to the highly integrated (i.e., highly intelligent) structure in Fig. \ref{fig:intelligent-transceiver}, ML-based transmission and processing systems can also be partially intelligentized. For example, in one scenario, only the channel coding or decoding block is managed by machine learning, meaning that the channel coding scheme is designed by machines rather than information scientists. In another scenario, machine learning is used solely for the transmit beamformer or the receive beamformer. A conceptual illustration is shown in Fig. \ref{fig:intelligent-modules}.

In technical details, applications of machine learning in physical layer communications include overall end-to-end system design \cite{ye2020deep,ye2021deep} (cf. Fig. \ref{fig:intelligent-transceiver}) and individual module design (cf. Fig. \ref{fig:intelligent-modules}). The latter, to be specific, encompasses 
\begin{itemize}
    \item coding/decoding techniques, for example, source coding \cite{manouchehri2022delay}, channel coding\cite{ye2019circular,zhang2021design}, and joint source-channel coding \cite{jankowski2020deep,yang2021deep}, 

    \item signal modulation and detection \cite{ye2017power,he2020model}, 
    
    \item transmit and receive beamforming \cite{saeed2022robust,ahmet2022family,brilhante2023literature,wang2024distributionally,shlezinger2024artificial}, for example, beam alignment and beam tracking \cite{lim2021deep,sohrabi2022active,wei2022fast,yi2024beam}, 
    
    \item channel estimation and feedback \cite{hu2020deep,guo2020convolutional}, 
\end{itemize}
among many others. For comprehensive and recent surveys, see, e.g., \cite{qin2019deep,burak2022deep,ye2024artificial}.

Coding and decoding techniques are vital in digital communications, ensuring efficient and reliable data transmission. Recently, machine learning has been increasingly applied to enhance these techniques, encompassing source coding, channel coding, and joint source-channel coding. Traditionally, source coding (i.e., data compression) reduces data redundancy for efficient transmission and storage. ML techniques, such as neural-network-based autoencoders, have revolutionized this field. Autoencoders learn efficient representations of data by encoding it into a lower-dimensional space and then reconstructing it, achieving high compression rates with minimal loss of information \cite{yang2023introduction}. Channel coding adds redundancy to data to detect and correct transmission errors caused by noisy channels. ML models, particularly deep learning techniques, have been applied to develop novel error correction codes. For example, neural decoders have been designed to decode complex schemes like low-density parity-check (LDPC) \cite{han2022deep} and Turbo codes \cite{jiang2019turbo}, offering improved performance over traditional algorithms, especially in highly noisy environments. Joint source-channel coding (JSSC) integrates source and channel coding to optimize overall system performance. ML models, such as variational autoencoders (VAEs) \cite{choi2019neural}, convolutional neural networks (CNNs) \cite{bourtsoulatze2019deep}, and generative adversarial networks (GANs) \cite{erdemir2023generative}, are used to jointly learn the representation and error correction codes. These models can adapt to the characteristics of both the source and the channel, achieving better compression and error resilience than traditional methods. Overall, ML-based coding and decoding techniques represent a significant advancement in digital communications. By leveraging the predictive and adaptive capabilities of ML, these techniques enhance data compression, error correction, and overall transmission efficiency, paving the way for more robust and efficient communication systems.

Signal modulation and detection are fundamental processes in digital communications, enabling the transmission and interpretation of data over various channels. Recently, machine learning techniques have been applied to enhance these processes, improving efficiency and reliability. Modulation involves altering a carrier signal's properties, such as amplitude, frequency, or phase, to encode information. Traditional modulation schemes include amplitude modulation (AM), frequency modulation (FM), and phase shift keying (PSK). ML techniques, particularly deep learning models, are now used to design adaptive modulation schemes. These models can dynamically adjust modulation parameters based on the channel conditions, optimizing performance in real time. For instance, neural networks can learn complex modulation patterns that maximize data throughput and minimize error rates \cite{bobrov2021massive}. Detection involves demodulating the received signal to recover the transmitted information. Traditional methods rely on predefined algorithms to estimate the transmitted data, often assuming specific channel characteristics. ML approaches, such as fully connected deep neural networks \cite{ye2017power} and transfer learning \cite{van2022transfer}, have been employed to enhance signal detection. These models can learn from data to accurately detect signals under varying and complex channel conditions, improving robustness against noise and interference. Overall, the integration of machine learning techniques in signal modulation and detection represents a significant leap forward in communication technology, which enhances data transmission efficiency, resilience to noise, and overall system performance.

Transmit and receive beamforming techniques are essential in wireless communication systems for enhancing signal quality and increasing data throughput. Beamforming directs the transmission or reception of signals in specific directions using antenna arrays, improving signal strength and reducing interference. Recently, machine learning techniques have significantly advanced beamforming performance and adaptability. 
For transmit beamforming, traditional methods, such as phased array systems, use predefined algorithms to adjust the phase and amplitude of signals from multiple antennas. In contrast, ML techniques, particularly deep learning models, optimize this process by learning from environmental data. As an example, reinforcement learning can dynamically adjust beamforming patterns in real time based on feedback from the communication environment, enhancing performance in complex and changing scenarios \cite{mismar2019deep,chu2022deep}. 
For receive beamforming, conventional methods, such as minimum variance distortionless response (MVDR) and maximal ratio combining (MRC), rely on statistical models of the signal environment. ML approaches, such as convolutional neural networks, improve upon these by learning optimal beamforming weights directly from data, allowing for more accurate and robust signal reception in diverse and dynamic environments \cite{huang2019fast,ramezanpour2020deep}. 
Beam alignment and tracking are crucial subcategories of beamforming, particularly important in high-frequency bands like millimeter-wave (mmWave) and terahertz communications. These techniques ensure that the transmitter and receiver maintain optimal beam alignment to maximize signal strength and data throughput \cite{wei2022fast,chen2023beam,yi2024beam}. Traditional alignment methods rely on exhaustive search or iterative algorithms, which are time-consuming and computationally intensive. ML approaches, such as supervised learning, multi-armed bandits, and reinforcement learning, provide more efficient solutions by predicting optimal beam directions from historical data, significantly reducing the search space. Beam tracking maintains alignment as the transmitter or receiver moves or as the environment changes. ML techniques, particularly deep learning models, enhance tracking by predicting beam direction changes in real time. Recurrent neural networks and long short-term memory networks, which capture temporal dependencies, are particularly effective for this purpose. For technical details on ML-based beam alignment and tracking, see, e.g., \cite{lim2021deep,sohrabi2022active,wei2022fast,chen2023beam,yi2024beam}. 
In summary, the integration of machine learning into beamforming, including beam alignment and tracking, is critical for next-generation networks such as 5G and beyond, because these ML-driven techniques can leverage predictive and adaptive capabilities to enhance signal quality, reduce interference, and optimize system performance.

Channel estimation and feedback techniques are of high importance in wireless communication systems for accurately characterizing the communication channel and ensuring efficient data transmission. These processes involve measuring the channel's properties and providing necessary feedback to transmitters. Recently, machine learning techniques have been applied to enhance these processes, offering significant improvements in accuracy and efficiency \cite{guo2022overview,guo2024deep}. Channel estimation involves predicting the state of the communication channel to optimize signal transmission and reception. Traditional methods, such as minimum mean-squared error (MMSE), rely on statistical models and require significant computational resources. Machine learning approaches, especially deep learning models, have introduced new ways to perform channel estimation with higher accuracy, and potentially, lower computational complexity. For instance, convolutional neural networks can learn to estimate channel states directly from received signal data, providing more robust and adaptive solutions in complex environments \cite{jiang2021dual}. Long short-term memory networks are particularly effective for capturing temporal dependencies in channel conditions, improving estimation accuracy \cite{shankar2023bi}. Feedback mechanisms forward channel state information (CSI) from the receiver back to the transmitter, allowing for real-time adaptation of transmission setups. Traditional feedback methods often involve quantizing and encoding the CSI, which can introduce delays and inaccuracies. Machine learning techniques, such as autoencoders and convolutional neural networks, improve feedback efficiency by compressing and reconstructing CSI with minimal loss of information \cite{guo2020convolutional}. This allows for more precise and timely adjustments to transmission strategies. In addition, ML models can simultaneously handle channel estimation and feedback, optimizing both processes in an integrated manner \cite{guo2024deep}. This holistic approach leverages the strengths of ML to enhance overall system performance.

\section{Semantic Communications}\label{sec:SematicCommu}
Semantic communications, unlike conventional physical-layer communications, focus on transmitting semantic information conveyed in original data (e.g., image, text, audio) rather than bit-wise raw information. The primary benefit of semantic communications is that the transmission overloads of wireless channels can be significantly reduced compared to bit-wise transmission. Consequently, the information transmission speed and efficiency can be considerably improved. For comprehensive and recent surveys in semantic communications, see, e.g., \cite{yang2022semantic-survey,luo2022semantic,lu2023semantics,chaccour2024less}.

\begin{figure}[!htbp]
    \centering
    \includegraphics[height=2.9cm]{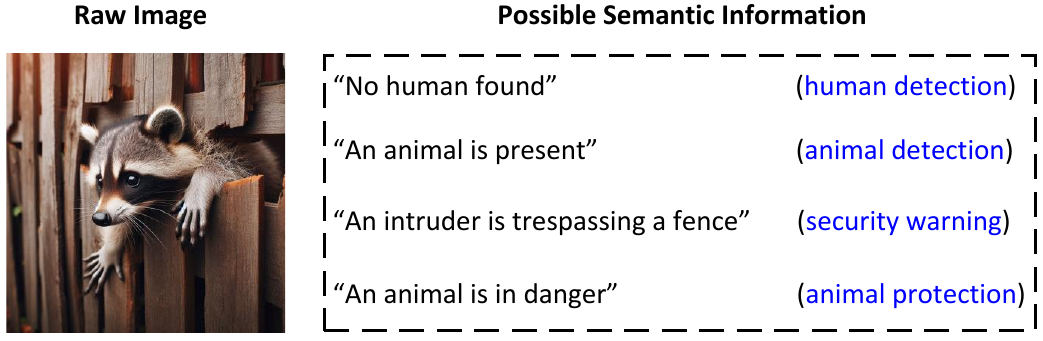}
    \caption{The semantic information of raw data is task-specific. Losslessly transmitting a high-definite image is time- and resource-consuming. However, accurately transmitting a semantic message can be relatively simpler and cheaper. (The image is generated by Microsoft Copilot.)}
    \label{fig:semantic-information}
\end{figure}

The key to semantic communications is to extract the semantic information from raw data. Therefore, semantic communications can be realized by elegantly designing the source coding and decoding strategies. It can also be actualized through joint source-channel coding and decoding. The difficulty, however, is that the semantic information of given raw data is specific to a task (see Fig. \ref{fig:semantic-information}), due to which, a generally well-accepted mathematical analysis, modeling, and computing framework for semantic communications is still lacking; for exploring works in this direction, see, e.g., \cite{gunduz2022beyond}. Therefore, for a specified communication task, the semantic coding and decoding schemes need to be elaborated. In the context of intelligent transmission and processing, semantic communications can be implicitly realized in highly integrated end-to-end transceivers; cf. Fig. \ref{fig:intelligent-transceiver}.

Machine learning approaches play a pivotal role in semantic communications, enabling systems to understand, process, and convey meaning more accurately. Techniques such as deep learning models, including transformers, convolutional neural networks, and recurrent neural networks, have been widely utilized to analyze and predict the semantic relevance of data, thus optimizing bandwidth usage and improving communication efficiency. To be specific, natural language processing (NLP) algorithms allow for the extraction and interpretation of semantic content from text and audio, facilitating more meaningful data compression and transmission \cite{chowdhary2020natural,khurana2023natural}; computer vision methods, on the other hand, enable the extraction and interpretation of semantic meaning from image and video \cite{szeliski2022computer}.

Recent research has demonstrated the potential of ML-driven semantic communications in various applications. For example, in \cite{xie2021deep}, transceiver neural networks have been designed to directly transmit text semantic meaning, which significantly reduces the demand on communication resources and improves the overall transmission performance; in \cite{jiang2022wireless}, an efficient system for video conferencing is developed to improve transmission efficiency. Most studies in semantic communications focus on joint semantic source and channel coding to save communication resources, which, however, requires changing the existing communication infrastructures and hinders practical implementation. As such, a pragmatic approach to wireless semantic transmission through revising some modules in existing infrastructures is reported \cite{jiang2023wireless}. To guarantee semantic transmission reliability and communication efficiency,  the spectral efficiency in the semantic domain and the semantic-aware resource allocation issues have been investigated in \cite{yan2022resource}. In addition to the above representative applications, the synergy between semantic communications and emerging technologies, such as the internet of things (IoT) \cite{xie2020lite} and edge computing \cite{yang2022semantic}, is fostering new opportunities for intelligent and context-aware communication systems. By leveraging distributed ML models, semantic communication systems can dynamically adapt to changing environmental conditions and user requirements, ensuring robust and efficient information exchange \cite{tong2021federated}. 

In summary, semantic communications, underpinned by advanced machine learning techniques, define a brilliant future direction for communication systems. This innovative approach promises to reform how information is transmitted and understood, offering profound implications for the efficiency and effectiveness of future communication networks.

\section{Resource Allocation in Communications}\label{sec:ResAllo}
\begin{figure}[!htbp]
    \centering
    \includegraphics[height=5.5cm]{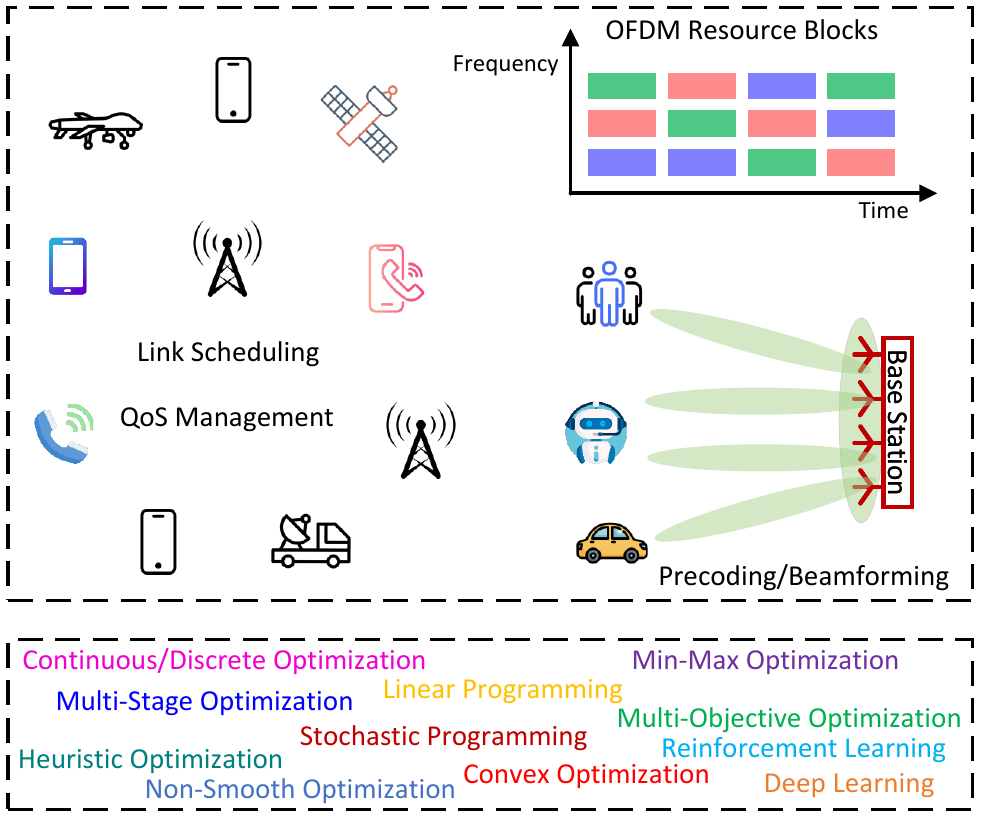}
    \caption{Typical applications and solution frameworks of resource allocation in wireless communications. OFDM: orthogonal frequency-division multiplexing; QoS: quality of service. (Icon Credit: FLATICON.com.)}
    \label{fig:resource-allocation}
\end{figure}

Resource allocation in wireless communications is concerned with how to efficiently manage and utilize the spectra, power, computing, space, and time resources, thus improving the overall communication network performances, e.g., higher throughput, lower latency, larger coverage, higher reliability, to name a few \cite{han2008resource,teng2018resource,liang2019deep}. Typical applications encompass link scheduling, message routing, power allocation, channel selection, beamforming, spectra access and management, and division protocol design (viz., time division, frequency division, etc.), among many others. From the mathematical programming perspective, resource allocation is often formulated as optimization problems. From the operations research perspective, assignment and scheduling are two pivotal techniques; the former handles static resource allocation problems, while the latter addresses dynamic ones; the static and dynamic features are with respect to time. From the computational and algorithmic perspective, standard and trending solution frameworks include the following:
\begin{itemize}
    \item continuous optimization, discrete (e.g., combinatorial, integer) optimization, and mixed optimization, 
    \item single-objective optimization and multi-objective optimization, 
    \item linear programming and nonlinear programming,
    \item convex optimization and non-convex optimization,
    \item smooth optimization and non-smooth optimization,
    \item min-max optimization (e.g., game theory, worst-case robust analyses),
    \item deterministic programming and stochastic programming (i.e., whether random variables are involved; if involved, associated distributions are considered),
    \item single-stage optimization (i.e., static programming) and multi-stage optimization (i.e., dynamic programming), 
    \item heuristic optimization (e.g., genetic algorithm, particle swarm optimization, simulated annealing),
    \item surrogate optimization which is also known as black-box optimization (e.g., Bayesian optimization), and 
    \item ML-based optimization (e.g., solution methods based on reinforcement learning and deep learning).
\end{itemize}
The canonical applications and solution frameworks of resource allocation in wireless communications are shown in Fig. \ref{fig:resource-allocation}. For introductory and motivational reading on this topic, refer to, e.g., \cite{han2008resource,zheng2016sequential,hossain2017radio}. 

Nowadays, the advent of integrated sensing and communication \cite{liu2022integrated} has slightly changed the connotation of traditional resource allocation. This is because the best resource allocation scheme for communication is not necessarily the same as (or in accordance with) that for sensing; see, e.g., \cite{dong2022sensing}. For example, the optimal waveforms for communication and sensing are usually dissimilar \cite{liu2022survey,wang2024robust} because the two radio functions have different or even contradicting design preferences. Therefore, diverse resources, including radio, computing, power, time, beam, etc., should be delicately allocated to satisfy the performance requirements of communication and sensing, respectively. The same dilemma holds for integrated computing and communication, e.g., edge computing \cite{mao2017survey} and networked control \cite{ge2017distributed} (NB: controllers are in nature information processors and therefore ad-hoc computing modules), because limited resources need to be elegantly distributed to computing and communication, respectively; refer to, e.g., \cite{luo2021resource,djigal2022machine}.

Machine learning techniques have emerged as powerful tools to address the challenges brought by resource allocation in wireless communications. Recent advancements have demonstrated the potential of machine learning in various resource allocation tasks. For instance, deep reinforcement learning has been applied to optimize spectrum allocation, power control, and user association in heterogeneous networks, showing significant improvements over conventional methods \cite{xu2017deep,liang2019spectrum,ye2019deep,zhao2019deep,sun2019application,xiong2020resource,xu2023distributed}. Similarly, supervised learning algorithms have been used to efficiently solve complex optimization problems in resource allocation such as mixed integer nonlinear programming (MINLP) \cite{lee2019learning}. In addition, unsupervised learning techniques can also be employed to solve resource allocation problems and refine the solutions, e.g., the graph embedding trick in link scheduling \cite{lee2020graph}. 

\begin{figure}[!htbp]
    \centering
    \includegraphics[width=8.8cm]{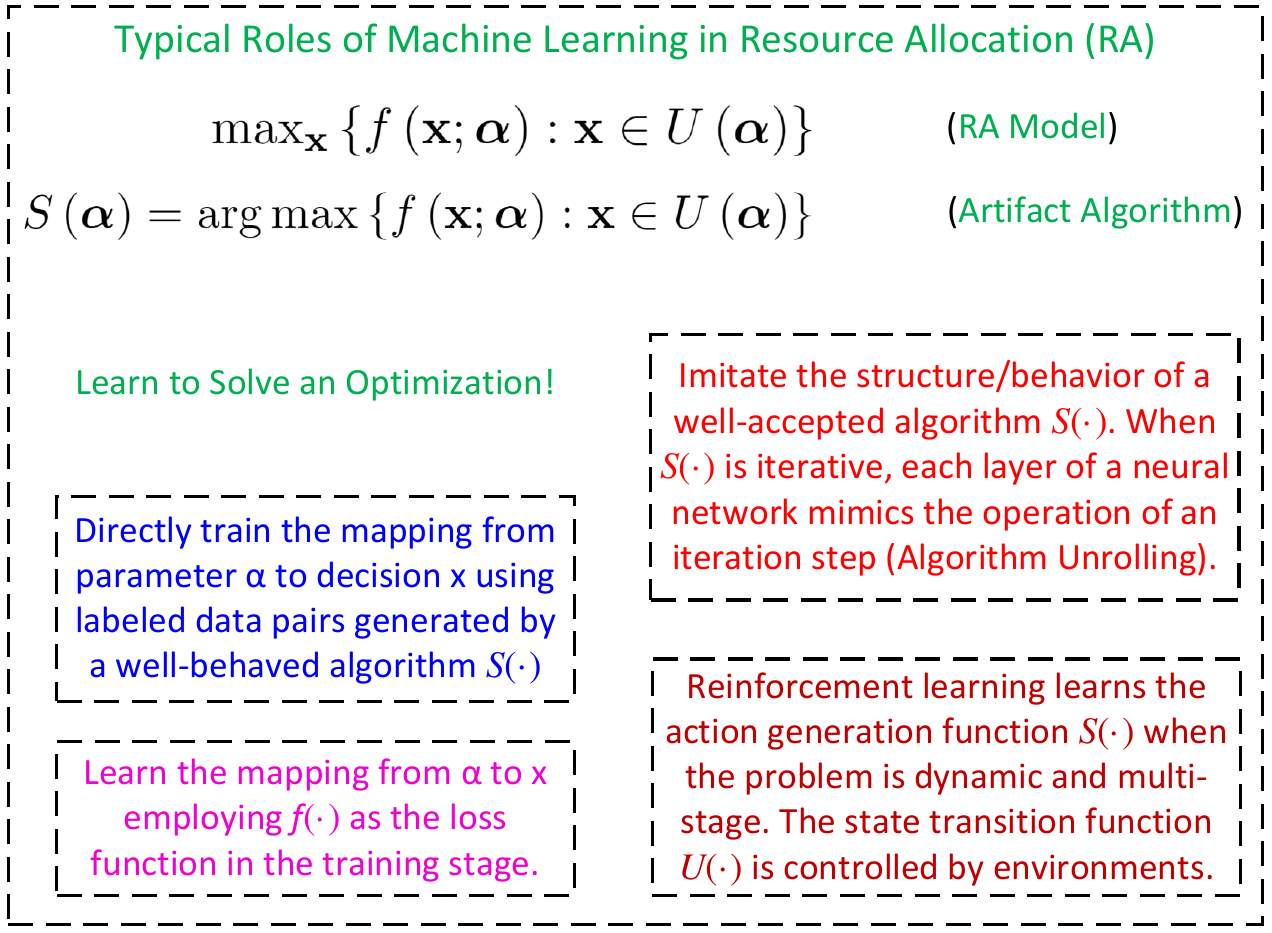}
    \caption{Four typical roles of machine learning in resource allocation---learning to solve optimizations.}
    \label{fig:ML-resource-allocation}
\end{figure}

Traditional resource allocation methods rely heavily on human intellect to build exact models and develop ad-hoc solution methods, which can be suboptimal and even inflexible in dynamic, complex, and large-scale scenarios. Machine learning, particularly through techniques like deep learning and reinforcement learning, offers the ability to model complex interactions with environments, predict future communication-network states, and optimize decisions in a real-time manner, thereby enhancing the overall performance and adaptability of wireless networks \cite{liang2019deep}. To be specific, for example, communication channels in practice are often time-varying, however, mathematically considering such model uncertainties is not straightforward. This is because we do not exactly know how the channels evolve over time. Even worse, the resultant mathematical programming models are computationally complex, and therefore, hard to be efficiently and optimally solved. The role of machine learning, in this sense, is to leverage accessible real-world data, discover the hidden knowledge and patterns that the data convey, and automatically find satisfactory resource allocation decisions. In technical details, on the one hand, machine learning can assist in solving computationally difficult optimization problems because resource allocation optimization can be seen as a mapping from parameters to decisions. This data-to-decision correspondence benefits from the powerful function-fitting ability of supervised learning based on deep neural networks, where labeled data-decision pairs are generated by well-behaved artifact solution methods. On the other hand, machine learning can treat the utility function of a resource allocation problem as the loss function in the training stage. This strategy allows machine learning to generate high-quality resource allocation decisions without relying on legacy human-made algorithms. In addition to the above two machine-learning schemes in resource allocation, another archetype, called algorithm unrolling \cite{monga2021algorithm}, employs neural networks to unroll existing efficient iterative algorithms. Specifically, each neural network layer acts as an iteration step of an iterative algorithm. By cascading several layers, an iteration process of the algorithm can be mimicked. This algorithm-instructed archetype is also referred to as model-driven deep learning \cite{he2019model}, where the architecture of a deep neural network is tailored considering domain knowledge, thus improving the generalizability of the network and reducing the required size of the training data set. The fourth promising machine learning paradigm in resource allocation is to use reinforcement learning to explore unknown and hard-to-model environments (e.g., complex and dynamic physical transmission channels). By interacting with environments, intelligent resource allocation solutions can be learned. The four typical roles of machine learning in resource allocation are summarized in Fig. \ref{fig:ML-resource-allocation}. The first benefit of using ML methods is their fast computing speed in the running stage, although the training state might be computationally heavy; cf. the first three schemes. The second benefit of using ML methods is the ability to respond to dynamic and uncertain (even unknown) environments without explicit physical modeling; cf. the fourth scheme, i.e., the reinforcement learning scheme.

\section{Beyond Data Transmission: Sensing and Computing}\label{sec:Beyond}
Wireless communication systems are undergoing transformative changes driven by the increasing demand for low-latency and high-speed connectivity, the growing need for sensing ability (e.g., to localize and track users) assisting high-performance communications, and the proliferation of connected devices enabling collaborative computing. This evolution has led to the development of innovative system paradigms such as integrated sensing and communications (ISAC) \cite{liu2020joint,liu2022integrated} and integrated computing and communications (ICAC) \cite{xu2023edge,wen2024integrated}, which aim to unify traditionally disparate functionalities to optimize resource usage, reduce hardware costs, and enhance overall system capabilities. For example, environmental and users' sensory data can be utilized to enhance communication performance through beam management and resource allocation, while sharing sensing data across network nodes enables real-time network monitoring and situational awareness for better sensing accuracy and larger coverage. For another example, local data processing at the edge can reduce latency for real-time communications, while high-speed communications enable efficient distributed computing for large-scale data analytics. As elucidated in previous sections, in modern communication systems, machine learning (especially deep learning) techniques are indispensable. They offer sophisticated algorithms that can learn from vast amounts of data, and therefore, optimize various aspects of communication networks, including resource allocation, signal processing, and fault detection. These benefits are also applicable to emerging ISAC and ICAC systems. In ISAC, deep learning models, such as convolutional neural networks and transformers, can improve sensing accuracy and robustness \cite{zhang2024white}, while realizing semantic information transmission \cite{zhang2024compression}. In ICAC, machine learning algorithms, such as federated learning, can protect users' data privacy and optimize computational tasks, facilitating efficient data processing and communication \cite{zhou2023fedgia,zhou2023federated}. In short, the synergy between ML/DL and the developing integrated paradigms enables more intelligent, adaptive, and efficient communication systems.

\subsection{Integrated Sensing and Communications}
ISAC is a paradigm that merges sensing and communication functionalities into a single system, leveraging shared infrastructure and spectral resources. This integration is essential in applications where both capabilities are crucial, such as autonomous vehicles, smart cities, and advanced surveillance systems. ISAC enhances the efficiency and performance of these systems by enabling simultaneous data acquisition and communication, thus reducing hardware costs and spectral congestion. This integration complicates the design of communication waveforms, the allocation of system and hardware resources, interference management, and overall network operations \cite[Table~1]{demirhan2023integrated}, \cite[Table~1]{lu2024integrated}. These challenges drive the need for novel approaches to unlock the potential of ISAC systems in real-world applications. Machine learning and deep learning techniques are, therefore, pivotal in ISAC, providing advanced data processing and decision-making capabilities. For comprehensive and motivational surveys on ML for ISAC, refer to, e.g., \cite{demirhan2023integrated,lu2024integrated}.

\subsection{Integrated Computing and Communications}
ICAC represents the convergence of computing and communication functionalities, aiming to meet the increasing computational demands of modern applications while maintaining high and robust communication performance. This integration is driven by the necessity to handle massive data processing tasks close to the source, reducing latency and improving efficiency of communications in edge computing environments, and enabling intelligence of all connected devices. ICAC facilitates real-time data processing and analytics, essential for applications like industrial automation, virtual reality, and the internet of things. Typical examples of ICAC include edge computing, federated learning, pervasive computing, fog computing, internet of things/vehicles, and autonomous systems, to name a few. Machine learning and deep learning are integral to ICAC, enabling dynamic resource allocation, adaptive system configurations, and real-time information analytics. These techniques ensure that computing and communication resources are utilized optimally, providing enhanced performance and responsiveness. For comprehensive and motivational surveys on ML for ICAC, refer to, e.g., \cite {alsamhi2022computing,ferrag2023edge,nugroho2024survey}. Note that, swarm intelligence and network control \cite{ge2017distributed} are highly related to ICAC because controllers are, in nature, information processors (mapping the system's state signals to the system's control input signals), and therefore, ad-hoc computing modules.

\section{Discussions and Conclusions}\label{sec:conclusion}
\begin{figure}[!htbp]
    \centering
    \includegraphics[width=7cm]{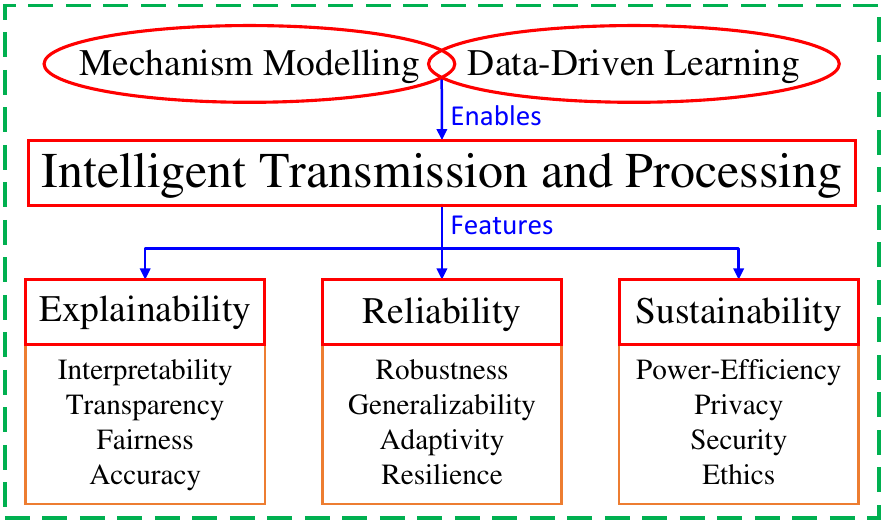}
    \caption{Features, challenges, and considerations of intelligent transmission and processing; cf. Fig. \ref{fig:nothing-free}. Although data-driven machine learning is powerful, mechanism modeling (including discovering physical/mathematical laws) is always important to improve explainability, reliability, and sustainability.}
    \label{fig:ITP-overview}
\end{figure}

This article discusses several pivotal aspects where machine learning can reform wireless communications, including but not limited to physical-layer communications, semantic communications, resource allocation, integrated sensing and communications, and integrated computing and communications (e.g., federated learning, edge computing). These applications demonstrate ML's potential to upgrade various facets of communication systems, ranging from signal processing algorithms to overall network management. Nevertheless, the adoption of ML in communications is not without challenges and its role should not be overstated. Issues, such as the interpretability and troubleshooting of ML models, the need for large and rich training datasets, and the high computational resources (e.g., power, processing speed) required for training and deployment, must be addressed. In addition, the reliability and security of ML-based systems should also be emphasized, particularly in scenarios where data privacy (cf. federated learning \cite{ye2022decentralized}), data freshness (cf. few-shot learning \cite{wang2023few,wang2022learn}), and real-time decision-making (e.g., autonomous driving) are critical. To address these challenges, the hybrid methodology that combines the strengths of traditional physical-law models with emerging data-driven ML models is advocated. Such a synergistic strategy can leverage the reliability and interpretability of physical mechanisms while harnessing the adaptability and learning capabilities of ML, thus enhancing overall communication system performance; see Fig. \ref{fig:ITP-overview} for features, challenges, and future considerations of intelligent transmission and processing. Among all the challenges that we can imagine, the following three items are of the crux in real-world operations because they are the minimum requirements to actualize an ML-based communication system:
\begin{itemize}
    \item How to interpret the performance gains and failures of machine-learned models, and how to troubleshoot and repair when systems are down, thus improving the overall reliability of systems? In this sense, the paradigm in Fig. \ref{fig:intelligent-modules} is more reliable and manageable than that in Fig. \ref{fig:intelligent-transceiver}.
    
    \item How to use practically limited data for better generalization and how to integrate newly available data to improve generalization capability \cite{wang2024distributional,wang2024distributionally}? This consideration also includes how to quickly adapt the learned model to new data, for example, when the environment's data-generating laws change over time \cite{wang2022learn,van2022transfer}. In machine learning terminologies, data freshness, sample efficiency, and data-distributional robustness are highly related to this issue.
    
    \item How to build domain-knowledge-informed machine learning models (beyond general-purpose deep neural networks such as multi-layer perception) and design computationally efficient training algorithms (beyond popular stochastic gradient descent) to diminish response times and power consumption \cite{he2019model,zhang2024white,liu2024energy,wang2024badm}? In addition, how to reduce the model sizes (especially those of deep neural networks) to save storage \cite{cai2020tinytl}? The three considerations above are particularly vital for embedded and edge devices.
\end{itemize}
In summary, the convergence of machine learning and communication systems marks a significant technological advancement, which offers the possibility for more intelligent, efficient, and reliable communication networks.

% \section{Acknowledgement}
% ChatGPT-4.0 was used to polish the language of this paper.

\bibliographystyle{IEEEtran}
\bibliography{References}

 \begin{IEEEbiography}[{\includegraphics[width=1in,height=1.25in,clip,keepaspectratio]{wsx}}]{Shixiong Wang} received the B.Eng. degree in detection, guidance, and control technology, and the M.Eng. degree in systems and control engineering from the School of Electronics and Information, Northwestern Polytechnical University, China, in 2016 and 2018, respectively. He received his Ph.D. degree from the Department of Industrial Systems Engineering and Management, National University of Singapore, Singapore, in 2022. 

 He is currently a Postdoctoral Research Associate with the Intelligent Transmission and Processing Laboratory, Imperial College London, London, United Kingdom, from May 2023. He was a Postdoctoral Research Fellow with the Institute of Data Science, National University of Singapore, Singapore, from March 2022 to March 2023.

 His research interest includes statistics and optimization theories with applications in signal processing (e.g., optimal estimation theory), machine learning (e.g., generalization error theory), and control technology.
 \end{IEEEbiography}

 \begin{IEEEbiography}
 [{\includegraphics[width=1in,height=1.2in,clip,keepaspectratio]{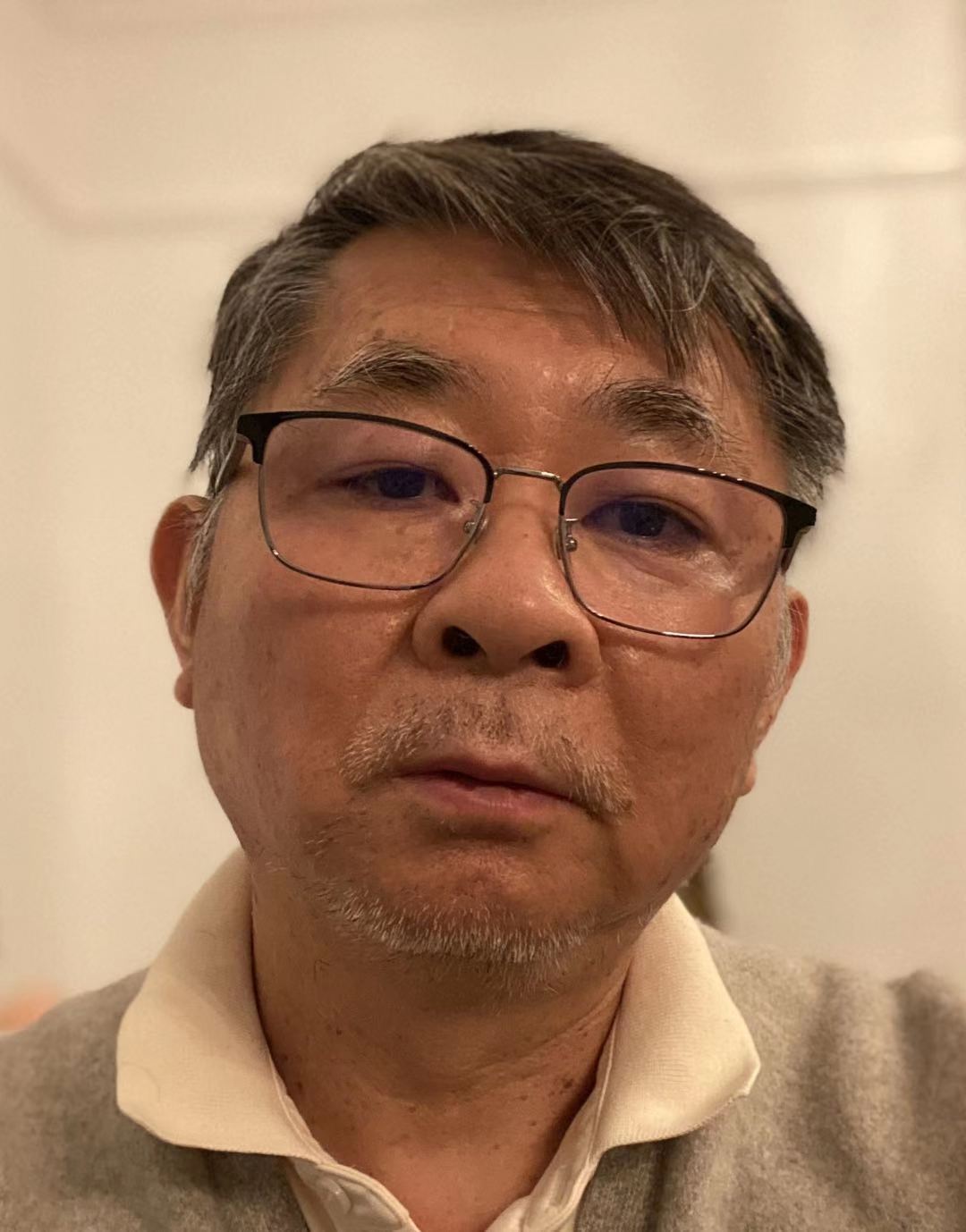}}] 
 {Geoffrey Ye Li} (Fellow, IEEE) is currently a Chair Professor at Imperial College London, UK. Before joining Imperial in 2020, he was a Professor at Georgia Institute of Technology, USA, for 20 years and a Principal Technical Staff Member with AT\&T Labs – Research (previous Bell Labs) in New Jersey, USA, for five years. He made fundamental contributions to orthogonal frequency division multiplexing (OFDM) for wireless communications, established a framework on resource cooperation in wireless networks, and introduced deep learning to communications. In these areas, he has published around 700 journal and conference papers in addition to over 40 granted patents. His publications have been cited over 69,000 times with an H-index of 119. He has been listed as a Highly Cited Researcher by Clarivate/Web of Science almost every year.

 Dr. Geoffrey Ye Li was elected to IEEE Fellow and IET Fellow for his contributions to signal processing for wireless communications. He won 2024 IEEE Eric E. Sumner Award, 2019 IEEE ComSoc Edwin Howard Armstrong Achievement Award, and several other awards from IEEE Signal Processing, Vehicular Technology, and Communications Societies.
 \end{IEEEbiography}

\end{document}